# Thickness-dependent insulator-to-metal transition in epitaxial $RuO_2$ films

Anil Kumar Rajapitamahuni[1,a], Sreejith Nair[1], Zhifei Yang[1,2], Anusha Kamath Manjeshwar[1], Seung Gyo Jeong[1], William Nunn[1], and Bharat Jalan[1,a]

[1]Department of Chemical Engineering and Materials Science, University of Minnesota, Minneapolis, MN 55455, USA

[2]School of Physics and Astronomy, University of Minnesota, Minneapolis, MN 55455, USA

a: Corresponding authors: arajapita1@bnl.gov; bjalan@umn.edu

**Abstract**

Epitaxially grown RuO$_2$ films on TiO$_2$ (110) exhibit significant in-plane strain anisotropy, with a compressive strain of - 4.7% along the [001] crystalline direction and a tensile strain of +2.3% along [1$\bar{1}$0]. As the film thickness increases, anisotropic strain relaxation is expected. By fabricating Hall bar devices with current channels along two in-plane directions <001> and <1$\bar{1}$0>, we revealed anisotropic in-plane transport in RuO$_2$/TiO$_2$ (110) films grown via solid-source metal-organic molecular beam epitaxy approach. For film thicknesses ($t_{\text{film}}$) ≤ 3.6 nm, the resistivity along <001> exceeds that along <1$\bar{1}$0> direction at all temperatures. With further decrease in film thicknesses, we uncover a transition from metallic to insulating behavior at $t_{\text{film}}$ ≤ 2.1 nm. Our combined temperature- and magnetic field-dependent electrical transport measurements reveal that this transition from metallic to insulating behavior is driven by electron-electron interactions.

Rutile ruthenium dioxide ($RuO_2$) holds significant technological importance due to its exceptional properties. It possesses high electrical conductivity and exhibits excellent thermal and chemical stability. These attributes render $RuO_2$ invaluable in applications such as catalysis, serving as an electrode material in energy storage devices, and acting as diffusion barriers in microelectronic devices[1]. From the perspective of correlated physics and magnetism, $RuO_2$ stands out as a 4d transition metal oxide (TMO) characterized by comparable bandwidths and on-site Coulomb repulsion energies. This unique combination of properties makes it intriguing, yet its behavior is not fully understood[2].

$RuO_2$ was long believed to be a Pauli paramagnet[3]. However, recent findings from polarized neutron diffraction[4] and resonant x-ray diffraction[5] studies challenge this notion. They provide evidence of an itinerant antiferromagnetic (AFM) ground state with a Neel temperature exceeding 300 K, and the Neel vector predominantly aligned along the c-axis. The AFM metallic properties of $RuO_2$ open up exciting possibilities for novel spintronics applications, including spin splitting torque[6-8] and AFM tunnel junction devices[9].

Similarly, the carrier transport properties of $RuO_2$ were thought to exhibit normal metallic behavior governed by electron-phonon and electron-electron interactions[10-13]. However, recent studies suggest $RuO_2$ is a Dirac nodal line semi-metal with flat band surface states[14] and is theoretically proposed to host a novel crystal Hall effect[15] thereby reviving itself as a quantum material. Furthermore, superconductivity has been observed below 2 K in thin films of $RuO_2$ grown on $TiO_2$ (110) substrates and the $T_c$ depends on the film thickness[16, 17]. No superconductivity is observed when thin films were grown on $MgF_2$ (110) substrates with similar crystal orientation but with smaller lattice mismatch. These findings suggest that strain might play an important role in the electronic properties of $RuO_2$ films. Additionally, it is noteworthy that $RuO_2$/$TiO_2$ (110) displays pronounced in-plane strain anisotropy with a compressive strain of - 4.7% along the [001] crystalline direction and a tensile strain of + 2.3% along the [1$\bar{1}$0] direction. These strains are further expected to exhibit a high sensitivity to variations in film thickness. Therefore, a careful study of thickness-dependent carrier transport properties in thin films of rutile $RuO_2$ is essential to investigate the possible role of strain-relaxation on electrical transport, and therefore, to potentially advance its applications in thin film spintronics devices.

In this work, we systematically vary film thickness ($t_{film}$) of RuO$_2$ film grown directly on a TiO$_2$ (110) substrate to investigate their temperature-dependent magneto-transport properties, and anisotropic transport behavior along two in-plane crystallographic directions, [001] and [1$\bar{1}$0]. With decreasing $t_{film}$, we observe a transition from metallic to insulating-like electrical transport behavior. The temperature dependent carrier transport properties at large thicknesses are primarily governed by electron-phonon interactions. The transition from metallic to insulating state in ultrathin films coincides with upturn in resistivity at low temperatures which can be explained by localization effects. Magnetotransport studies reveal electron-electron interactions play an important role at low temperatures at these thicknesses and can be explained by the theory of weak-antilocalization. Insulating behavior was observed at $t_{film}$ < 2 nm and is explained by strong localization which is likely due to disorder or other unknown effects at the substrate/film interface.

Epitaxial RuO$_2$ films with atomically smooth surfaces on TiO$_2$ (110) substrates were grown using a solid-source metal-organic MBE technique[18, 19]. Details of growth conditions and structural characterization can be found in ref[20]. The (110)-oriented TiO$_2$ substrates provide anisotropic strain to the RuO$_2$ film, resulting in a larger out of plane lattice parameter in thin films of RuO$_2$ compared to the bulk value (3.176 Å)[21]. To understand the effect of anisotropic strain on the transport properties, Hall bars with current channels along <001> and <1$\bar{1}$0> crystallographic directions were fabricated as shown in Fig.1(a). The fabrication of Hall bars is carried out by defining photoresist masks followed by Argon ion milling. Electrical contacts are made by Al wire bonding. DC currents of 100 nA – 10 µA are used to measure the channel resistance, using a Keithley 2430 source measure unit. Temperature-dependent magnetotransport measurements were performed between 1.8 K and 300 K in a Quantum Design Dynacool PPMS equipped with a superconducting 9 T magnet.

We examine the influence of film thickness on the electrical resistivity ($\rho$) in relation to temperature ($T$) using Hall bar devices along both crystalline orientations as shown in Fig. 1(b). As $t_{film}$ decreases, the resistivity ($\rho(T)$) increases for both crystalline orientations. Notably, the electrical resistivity along the <1$\bar{1}$0> and <001> directions exhibit distinct values. For $t_{film} \geq 3.6$ nm, the resistivity along the <1$\bar{1}$0> direction ($\rho_{<1\bar{1}0>}$) surpasses that along the <001> direction ($\rho_{<001>}$) at all temperatures. However, as the film thickness drops below 3.6 nm, this trend reverses and is quantitatively illustrated in Fig. 1(b), where we plot the resistivity ratio ($\rho_{<001>}/\rho_{<1\bar{1}0>}$) for the two crystal directions as a function of $t_{film}$ at two different temperatures.

The dependence of resistivity on film thickness, showcasing this anisotropic behavior, implies the existence of a complex strain relaxation mechanism in rutile $RuO_2$ films grown on $TiO_2$ (110) substrates. Prior studies also support the thickness-driven strain relaxation in $RuO_2/TiO_2(110)$[16, 17, 20]. It is worth noting that anisotropic strain relaxation mechanisms are known to impact orbital occupancy[22] and induce phase separation[23] in rutile $VO_2$. Consequently, our future investigations should be directed towards a more comprehensive exploration of these aspects within $RuO_2$ films.

In this study, we delved into the mechanisms governing transport behavior in these films. By examining ρ *vs.* *T* data presented in Fig.1(c), we divided these films into three distinct thickness regimes: (1) thick films ($t_{film} \geq 3.6$ nm) where films remain metallic down to the lowest temperature (1.8 K) in our measurements. These films also possess ρ < 150 μΩ-cm at 300 K. (2) Intermediate thickness ($t_{film}$ = 2.6 nm, and 2.1 nm). In this regime, films exhibited a slight upturn in the resistivity at low temperature (marked by an arrow in Fig. 1b), and a value exceeding 150 μΩ-cm at 300 K. The threshold resistivity of ~150 μΩ-cm regime falls well under the Mooij criterion[24], suggesting that the upturn in the resistivity is due to carrier localization. (3) ultrathin films ($t_{film} \leq$ 1.7 nm) where the resistivity increases with decreasing *T* for all temperatures ≤ 300K. For $t_{film}$ = 1.7 nm, resistivity approaches nearly 5000 μΩ-cm at room temperature (equivalent to the quantum resistance, $h/e^2$ ~26 kΩ/□), suggesting a crossover from weak to strong localization.

We next turn to the discussion of the data for films within these three thickness regimes. Figure 2a shows ρ *vs.* *T* data for a 14 nm $RuO_2/TiO_2$ (110) exhibiting metallic behavior down to 1.8 K along both crystalline directions <1$\bar{1}$0> and <001>. The residual resistivity ($\rho_0$) value drops between 18 - 20 μΩ-cm at 1.8 K, comparable or better to the previously reported values[16, 17]. Insets show ρ *vs.* $T^2$ data with a linear fit (solid black line) for 1.8 K < T< 25 K suggesting the resistivity takes the form $\rho = \rho_0 + AT^n$, where the value of *n* gives information about the scattering mechanisms and *A* is the resistance prefactor for the corresponding scattering type[25]. The value of *n* = 2 for 1.8 K ≤ T ≤ 25 K, suggest a Fermi liquid behavior with electron-electron scattering as the dominating mechanism with A = 2 × $10^{-4}$ - 5 × $10^{-4}$ μΩ-cm $K^{-2}$ along both crystallographic directions. For moderate temperatures, 25 K < T ≤ 150 K, $\rho(T)$ deviates from the Fermi liquid behavior, with *n* < 2. For T ≥ 150 K, the value of *n* is 4/5 (see solid red line in Fig. 2a), with $\delta\rho/\delta T$ ~ 1 μΩ-cm K$^{-1}$. In contrast to the normal metallic behavior (where *n* = 1, implying $\rho \propto T$) as observed in bulk single crystals, the sublinear $\rho(T)$ hints towards modified phonon modes in our films which may likely be due to the presence of anisotropic strain.

We next examine the transport properties of the films in the intermediate thickness regime near metal to insulator transition. For the intermediate thickness regime, we first check the transport dimensionality by comparing the carrier mean free path to the film thickness. The carrier mean free path is given by $l = h/\rho n e^2 \lambda_F$, where $\lambda_F = 2(\pi/3n)^{1/3}$. Here *n* is carrier density, $\lambda_F$ is the Fermi wavelength, e is electron charge and $h$ is Planck's constant. For $n \sim 10^{23} cm^{-3}$, which are typical values in our films, $\lambda_F$ is ~ 5 Å. The calculated values of $l$ ~ 5 Å are much smaller than the individual film thicknesses studied in this work, suggesting the carrier transport in these films is three-dimensional (3D). The low temperature-dependence of conductivity in 3D disordered systems is given by[25]

$$\sigma = \sigma_0 + AT^{1/2} + BT^{p/2} \qquad (1)$$

The second term arises due to the electron-electron interactions (EEI) and the third term is the correction to zero temperature conductivity due to localization effects. A and B are prefactors. The temperature dependence of the localization effect is determined by the temperature dependence of the scattering rate $\tau_\phi^{-1} = T^p$ of the dominant dephasing mechanism[26]. Figure 2(b) shows conductivity as a function of T$^{1/2}$ for the 2.6 nm film. Excellent agreement is shown between the data and equation (1) (red solid line), suggesting the presence of both EEI and localization effects in our thin films. The value of *p* is 3 for electron-phonon scattering, while it is 2 and 3/2 for inelastic electron-electron collisions in the clean and dirty limit respectively[25]. The extracted p values lie between 1.5 and 2, suggesting the dephasing mechanism at low temperatures is mediated by inelastic electron-electron collisions but not electron-phonon.

While the scattering rate due to EEI is sensitive to temperature, the localization effects are mainly limited by magnetic fields as they are averaged quantum effects. The application of a small magnetic field can destroy the time reversal symmetry which then breaks the phase coherence between the self-intersecting scattering paths of electron partial waves scattered by impurities. Therefore, temperature dependent magnetoresistance (MR) measurements were performed to determine the dominant contribution to the low temperature resistivity upturn behavior in our films. Prior to discussing the data as presented in Fig. 3, we discuss the transport property of the film in ultra-thin regime ($t_{film}$ = 1.7 nm), where strong localization exists. Here, we expect the low temperature conduction to occur via carrier hopping between the localized states, governed by variable range hopping (VRH) type conduction. The temperature dependence of hopping conduction is given by equation 2[27]

$$\sigma = C \, exp[-(T_0/T)^m], \qquad (2)$$

where $T_0$ is related to the density of localized states at the Fermi level and $m = 1/(d + 1)$ for non-interacting $d$ dimensional system. Attempts to fit the logarithm of sheet conductance of the 1.7 nm film as a function of $1/T^m$ for a 3D system ($m = 1/4$) and for a dimensional crossover from a 3D metallic to a 2D insulating state ($m = 1/3$) were not successful. Finally, by considering the role of Coulomb interactions[28], where $m = 1/2$, an excellent agreement with linear dependence is achieved in the temperature range 250 K to 4.2 K, as shown in Fig. 2(c). The slope of the Zabrodski plot[27], a log-log plot of the reduced activation energy, $W = \frac{d \, ln\rho(T)}{d \, lnT}$ vs $T$, in the inset of Fig. 2(c) is ~ ½. This observation supports our hypothesis of the role of Coulomb interactions on the conduction in the insulating regime and that the carrier transport is governed by Efros-Shklovskii variable range hopping[28].

We now turn to the discussion of temperature dependent MR measurements in Fig. 3 to determine the dominant contribution to the low temperature resistivity upturn for the intermediate thickness range. Figure 3(a) shows $\Delta\rho/\rho$ as a function of applied magnetic field ($H$) in the temperature range where an upturn in resistivity is observed for the 2.1 nm thick. At high fields, we observe a classical Lorentzian contribution to the MR which is quadratic and positive at all temperatures. In contrast, the low-field MR is qualitatively different suggesting the presence of a phase relaxation length ($l_\phi$) parametrically longer than the scattering mean-free path to observe this effect. In addition, the positive (negative) correction to the MR implies strong (weak) spin-orbit interactions, $l_{SO} \ll l_\phi$ ($l_{SO} \gg l_\phi$), where $l_{SO}$ is the spin-orbit scattering length. This positive (negative) interference corrections to the low-field MR are known as weak antilocalization (WAL) (weak localization (WL)) effects and the coherence is destroyed by the application of a magnetic field H ≥ $H_\phi$ = $h/(8\pi e l_\phi^2)$. For films with thickness less than $l_\phi$, in the absence of SOC, the correction to the conductance given by 2D WAL theory[29] by the Hikami-Larkin-Nagaoka (HLN) model is

$$\Delta\sigma(H) \cong \alpha(e^2/\pi h)f(H_\phi/H) \qquad (3)$$

where f(z) ≡ ln z – ψ (1/2 + z), with ψ being the digamma function and the value of $\alpha$ = ½ and 1 for one and two independent conduction channels respectively[30]. By assuming a single conducting channel, equation (3) has only one fitting parameter $l_\phi$, which can be determined by fitting the MR data. $l_\phi$ decreases with increasing temperature and is proportional to $T^{-p/2}$, where the value of exponent $p$ depends on the dephasing mechanisms. For EEI, $p$ takes values of 0.66, 1 and 1.5 for 1D, 2D and 3D respectively, while it is 2 - 4 for electron-phonon interactions[31]. Figure 3(b,c,) shows the magnetoconductance data along the <1$\bar{1}$0> and <001> crystallographic directions for films of thickness ~ 2 nm. Excellent fits to equation (3) are achieved, allowing us to determine $l_\phi$ as a function of temperature. The extracted $l_\phi$ are greater than the film thicknesses for all temperatures, justifying our assumption to use 2D WAL. $l_\phi$ decreases with $T$ and from the log-log plot of $l_\phi$ vs T (Fig. 3 (d,e)), we determine the value of $p$ to be ~ 0.76, suggesting a quasi 1D EEI as the dephasing mechanism. The value of $p$ is substantially lower than the value expected for 3D electron systems with diffusive transport. Alternatively, we have also considered 3D WAL model[32] to fit our data in which case the MR would behave as $H^{1/2}$. In this case, from the temperature dependence of $l_\phi$, the exponent $p$ came to be ~ 0.66. Details about the fitting can be found in the supplementary text and Fig S1. However, here it needs to be considered that the films are close to the metal-insulator transition (MIT) and the transport is more like VRH rather than diffusive. In which case $p = 2/(d+1)$, where $d$ is the dimensionality of the system[33]. It follows $p$ is 2/3 and ½ for the VRH transport in 2D and 3D electron systems. These values are closer to the experimentally obtained values from the fittings to both the 2D and 3D WAL models. Notwithstanding the dimensionality of the dephasing mechanism, it is clearly established EEI is the dominant dephasing mechanism but not the electron-phonon scattering with a value of $p$ = 3. Finally, the assumption of a single transport channel needs to be revisited as there could be surface conduction channels through scattering and their coupling with the bulk needs to be considered.

In conclusion, we have observed anisotropic electron transport in ultrathin films of $RuO_2$. With the reduction in thickness of the film, a metal to insulator transition is observed at room temperature. We observe a VRH type conduction in the insulating state and is characterized by the presence of a Coulomb gap due to the interactions between the strongly localized electrons. The electrical transport studies from the intermediate thickness regime suggest, the transition to insulating behavior is facilitated by both EEI and weak anti-localization. A detailed investigation of the magneto-transport properties in this thickness regime suggests the dominant dephasing mechanism is through electron-electron collisions. We believe that our study shed insights into the understanding of electronic transport in ultrathin films of $RuO_2$ which will further add to the ongoing study on the origins of superconductivity, magnetic order, spin currents and their manipulation using epitaxial growth techniques.

**References:**

1. Over, H., Surface Chemistry of Ruthenium Dioxide in Heterogeneous Catalysis and Electrocatalysis: From Fundamental to Applied Research. *Chem Rev* **2012,** *112* (6), 3356-3426.
2. Kim, H. D.; Noh, H. J.; Kim, K. H.; Oh, S. J., Core-level X-ray photoemission satellites in ruthenates: A new mechanism revealing the Mott transition. *Phys Rev Lett* **2004,** *93* (12).
3. Ryden, W. D.; Lawson, A. W., Magnetic Susceptibility of Iro2 and Ruo2. *J Chem Phys* **1970,** *52* (12), 6058-&.
4. Berlijn, T.; Snijders, P. C.; Delaire, O.; Zhou, H. D.; Maier, T. A.; Cao, H. B.; Chi, S. X.; Matsuda, M.; Wang, Y.; Koehler, M. R.; Kent, P. R. C.; Weitering, H. H., Itinerant Antiferromagnetism in RuO2. *Phys Rev Lett* **2017,** *118* (7).
5. Zhu, Z. H.; Strempfer, J.; Rao, R. R.; Occhialini, C. A.; Pelliciari, J.; Choi, Y.; Kawaguchi, T.; You, H.; Mitchell, J. F.; Shao-Horn, Y.; Comin, R., Anomalous Antiferromagnetism in Metallic RuO2 Determined by Resonant X-ray Scattering. *Phys Rev Lett* **2019,** *122* (1).
6. Bai, H.; Han, L.; Feng, X. Y.; Zhou, Y. J.; Su, R. X.; Wang, Q.; Liao, L. Y.; Zhu, W. X.; Chen, X. Z.; Pan, F.; Fan, X. L.; Song, C., Observation of Spin Splitting Torque in a Collinear Antiferromagnet RuO2. *Phys Rev Lett* **2022,** *128* (19).
7. Bose, A.; Schreiber, N. J.; Jain, R.; Shao, D. F.; Nair, H. P.; Sun, J. X.; Zhang, X. S.; Muller, D. A.; Tsymbal, E. Y.; Schlom, D. G.; Ralph, D. C., Tilted spin current generated by the collinear antiferromagnet ruthenium dioxide. *Nat Electron* **2022,** *5* (5), 267-274.
8. Gonzalez-Hernandez, R.; Smejkal, L.; Vyborny, K.; Yahagi, Y.; Sinova, J.; Jungwirth, T.; Zelezny, J., Efficient Electrical Spin Splitter Based on Nonrelativistic Collinear Antiferromagnetism. *Phys Rev Lett* **2021,** *126* (12).
9. Shao, D. F.; Zhang, S. H.; Li, M.; Eom, C. B.; Tsymbal, E. Y., Spin-neutral currents for spintronics. *Nat Commun* **2021,** *12* (1).
10. Glassford, K. M.; Chelikowsky, J. R., Electron transport properties in ${\mathrm{RuO}}_{2}$ rutile. *Physical Review B* **1994,** *49* (11), 7107-7114.
11. Lin, J. J.; Huang, S. M.; Lin, Y. H.; Lee, T. C.; Liu, H.; Zhang, X. X.; Chen, R. S.; Huang, Y. S., Low temperature electrical transport properties of RuO2 and IrO2 single crystals. *J Phys-Condens Mat* **2004,** *16* (45), 8035-8041.
12. Ryden, W. D.; Lawson, A. W., Electrical Transport Properties of Iro2 and Ruo2. *Physical Review B* **1970,** *1* (4), 1494-&.
13. Ryden, W. D.; Lawson, A. W.; Sartain, C. C., Temperature Dependence of Resistivity of Ruo2 and Iro2. *Phys Lett A* **1968,** *A 26* (5), 209-&.
14. Sun, Y.; Zhang, Y.; Liu, C. X.; Felser, C.; Yan, B. H., Dirac nodal lines and induced spin Hall effect in metallic rutile oxides. *Physical Review B* **2017,** *95* (23).
15. Smejkal, L.; Gonzalez-Hernandez, R.; Jungwirth, T.; Sinova, J., Crystal time-reversal symmetry breaking and spontaneous Hall effect in collinear antiferromagnets. *Sci Adv* **2020,** *6* (23).
16. Ruf, J. P.; Paik, H.; Schreiber, N. J.; Nair, H. P.; Miao, L.; Kawasaki, J. K.; Nelson, J. N.; Faeth, B. D.; Lee, Y.; Goodge, B. H.; Pamuk, B.; Fennie, C. J.; Kourkoutis, L. F.; Schlom, D. G.; Shen, K. M., Strain-stabilized superconductivity. *Nat Commun* **2021,** *12* (1).
17. Uchida, M.; Nomoto, T.; Musashi, M.; Arita, R.; Kawasaki, M., Superconductivity in Uniquely Strained RuO2 Films. *Phys Rev Lett* **2020,** *125* (14).

18. Nunn, W.; Manjeshwar, A. K.; Yue, J.; Rajapitamahuni, A.; Truttmann, T. K.; Jalan, B., Novel synthesis approach for "stubborn" metals and metal oxides. *P Natl Acad Sci USA* **2021,** *118* (32).
19. Nair, S.; Yang, Z.; Lee, D.; Guo, S.; Sadowski, J. T.; Johnson, S.; Saboor, A.; Comes, R. B.; Jin, W.; Mkhoyan, K. A.; Janotti, A.; Jalan, B., Engineering Metal Oxidation using Epitaxial Strain. *Nat. Nanotechnol.* **2023,** *18*, 1005.
20. Nunn, W.; Nair, S.; Yun, H.; Manjeshwar, A. K.; Rajapitamahuni, A.; Lee, D.; Mkhoyan, K. A.; Jalan, B., Solid source metal-organic molecular beam epitaxy of epitaxial $RuO_2$. *APL Mater.* **2021,** *9*, 091112.
21. Nunn, W.; Nair, S.; Yun, H. H.; Manjeshwar, A. K.; Rajapitamahuni, A.; Lee, D. Y.; Mkhoyan, K. A.; Jalan, B., Solid-source metal-organic molecular beam epitaxy of epitaxial $RuO2$. *Apl Mater* **2021,** *9* (9).
22. Aetukuri, N. B.; Gray, A. X.; Drouard, M.; Cossale, M.; Gao, L.; Reid, A. H.; Kukreja, R.; Ohldag, H.; Jenkins, C. A.; Arenholz, E.; Roche, K. P.; Durr, H. A.; Samant, M. G.; Parkin, S. S. P., Control of the metal-insulator transition in vanadium dioxide by modifying orbital occupancy. *Nat Phys* **2013,** *9* (10), 661-666.
23. Liu, M. K.; Wagner, M.; Abreu, E.; Kittiwatanakul, S.; McLeod, A.; Fei, Z.; Goldflam, M.; Dai, S.; Fogler, M. M.; Lu, J.; Wolf, S. A.; Averitt, R. D.; Basov, D. N., Anisotropic Electronic State via Spontaneous Phase Separation in Strained Vanadium Dioxide Films. *Phys Rev Lett* **2013,** *111* (9).
24. Mooij, J. H., Electrical-Conduction in Concentrated Disordered Transition-Metal Alloys. *Phys Status Solidi A* **1973,** *17* (2), 521-530.
25. Lee, P. A.; Ramakrishnan, T. V., Disordered Electronic Systems. *Rev Mod Phys* **1985,** *57* (2), 287-337.
26. Lin, J. J.; Bird, J. P., Recent experimental studies of electron dephasing in metal and semiconductor mesoscopic structures. *J Phys-Condens Mat* **2002,** *14* (18), R501-R596.
27. Mott, N. F., Conduction in non-crystalline materials. *The Philosophical Magazine: A Journal of Theoretical Experimental and Applied Physics* **1969,** *19* (160), 835-852.
28. Efros, A. L.; Shklovskii, B. I., Coulomb Gap and Low-Temperature Conductivity of Disordered Systems. *J Phys C Solid State* **1975,** *8* (4), L49-L51.
29. Hikami, S.; Larkin, A. I.; Nagaoka, Y., Spin-Orbit Interaction and Magnetoresistance in the 2 Dimensional Random System. *Prog Theor Phys* **1980,** *63* (2), 707-710.
30. Garate, I.; Glazman, L., Weak localization and antilocalization in topological insulator thin films with coherent bulk-surface coupling. *Physical Review B* **2012,** *86* (3).
31. Altshuler, B. L.; Aronov, A. G.; Khmelnitsky, D. E., Effects of Electron-Electron Collisions with Small Energy Transfers on Quantum Localization. *J Phys C Solid State* **1982,** *15* (36), 7367-7386.
32. Kawabata, A., Theory of Negative Magnetoresistance .1. Application to Heavily Doped Semiconductors. *J Phys Soc Jpn* **1980,** *49* (2), 628-637.
33. Liao, J.; Ou, Y. B.; Liu, H. W.; He, K.; Ma, X. C.; Xue, Q. K.; Li, Y. Q., Enhanced electron dephasing in three-dimensional topological insulators. *Nat Commun* **2017,** *8*.

**Acknowledgments**

This work was supported primarily by the Air Force Office of Scientific Research (AFOSR) through Grant Nos. FA9550-21-1-0025, FA9550-21-0460 and FA9550-23-1-0247. Film growth was performed using instrumentation funded by AFOSR DURIP award FA9550-18-1-0294 and FA9550-23-1-0085. Parts of the work were supported partially by the UMN MRSEC program under Award No. DMR-2011401. Parts of this work were carried out at the Characterization Facility, University of Minnesota, which receives partial support from the NSF through the MRSEC program under award DMR-2011401. Device fabrication was carried out at the Minnesota Nano Center, which is supported by the NSF through the National Nano Coordinated Infrastructure under award ECCS-2025124.

**Author Contributions:** A.K.R. and B.J. conceived the idea and designed the experiments. S.N., A.K.M., S.G.J, and W.N. grew the films. A.K.R. and Z.Y. performed electrical testing. A.K.R., S.N., and B.J. wrote the manuscript. All authors contributed to the discussion and manuscript preparation. B.J. directed the overall aspects of the project.

**Competing Interest Statement:** The authors declare no competing interests.

**Data and materials availability:** All data needed to evaluate the conclusions of the paper are present in the paper and/or the Supplementary Materials.

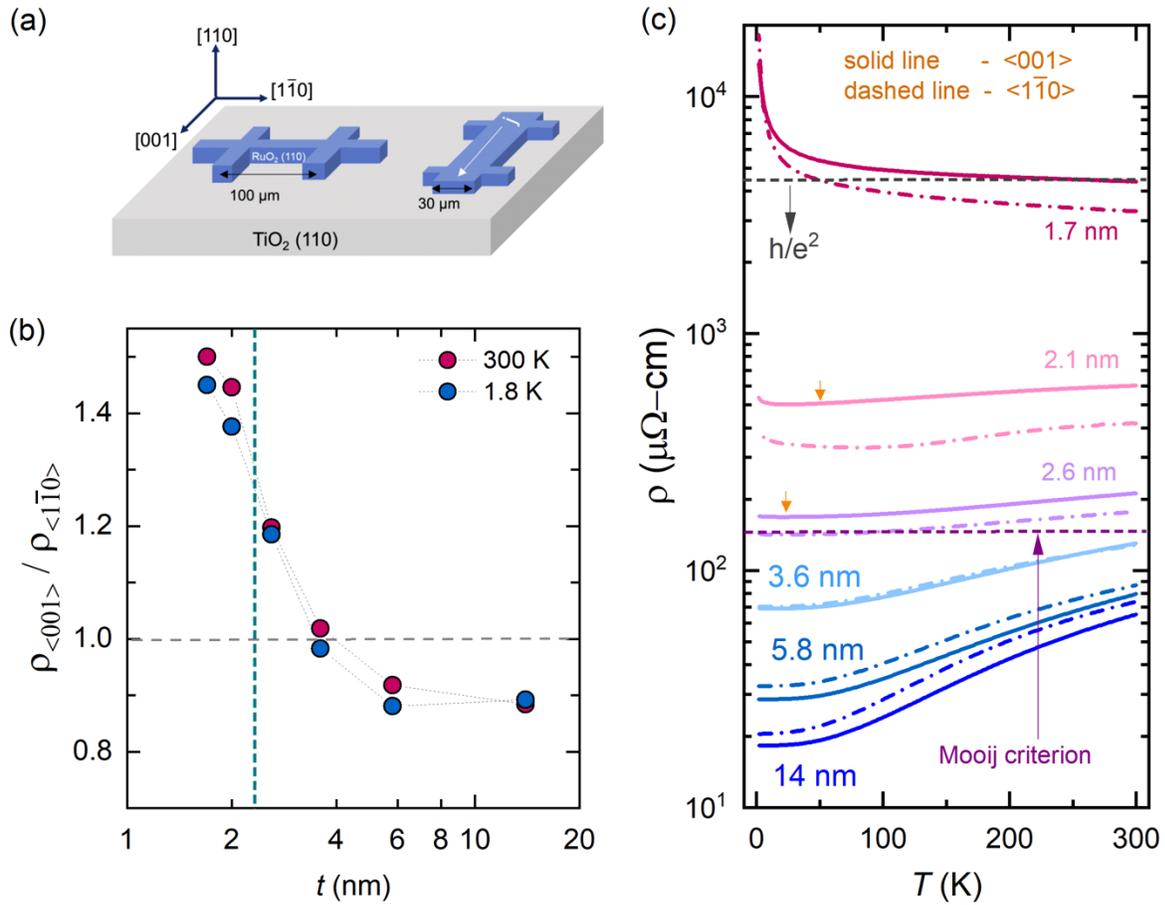

**Figure 1:** (a) Schematic of thin film Hall bar devices with current channels along the in-plane crystalline directions. (b) Ratio of resistivity along two orthogonal in-plane directions as a function of thickness ($t$) at 300 K and 1.8 K. The vertical dashed line separates the metallic and insulating regimes. (c) Resistivity vs temperature for different film thicknesses. The two orange arrows mark the temperature at which upturn in the resistivity occurs.

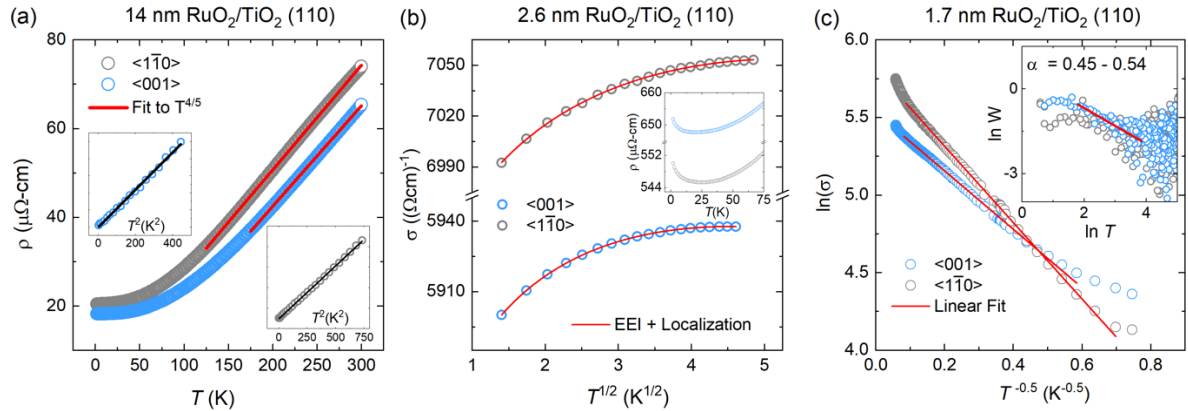

**Figure 2:** (a) Resistivity vs temperature for the metallic 14 nm film. The straight line corresponds to the fits for $T^{4/5}$ behavior. The insets show resistivity vs $T^2$ for temperatures below 25 K. Solid black line in the inset shows the linear fit to the data. (b) Conductivity as a function of $T^{1/2}$ for 2.6 nm film. The fits are the correction to conductivity using electron-electron interaction and localization effect. The inset shows low temperature upturn in resistivity on a linear scale. (c) Logarithm of conductivity as a function of $T^{-1/2}$ for the 1.7 nm film. The inset shows Zabrodski plot with slope ½.

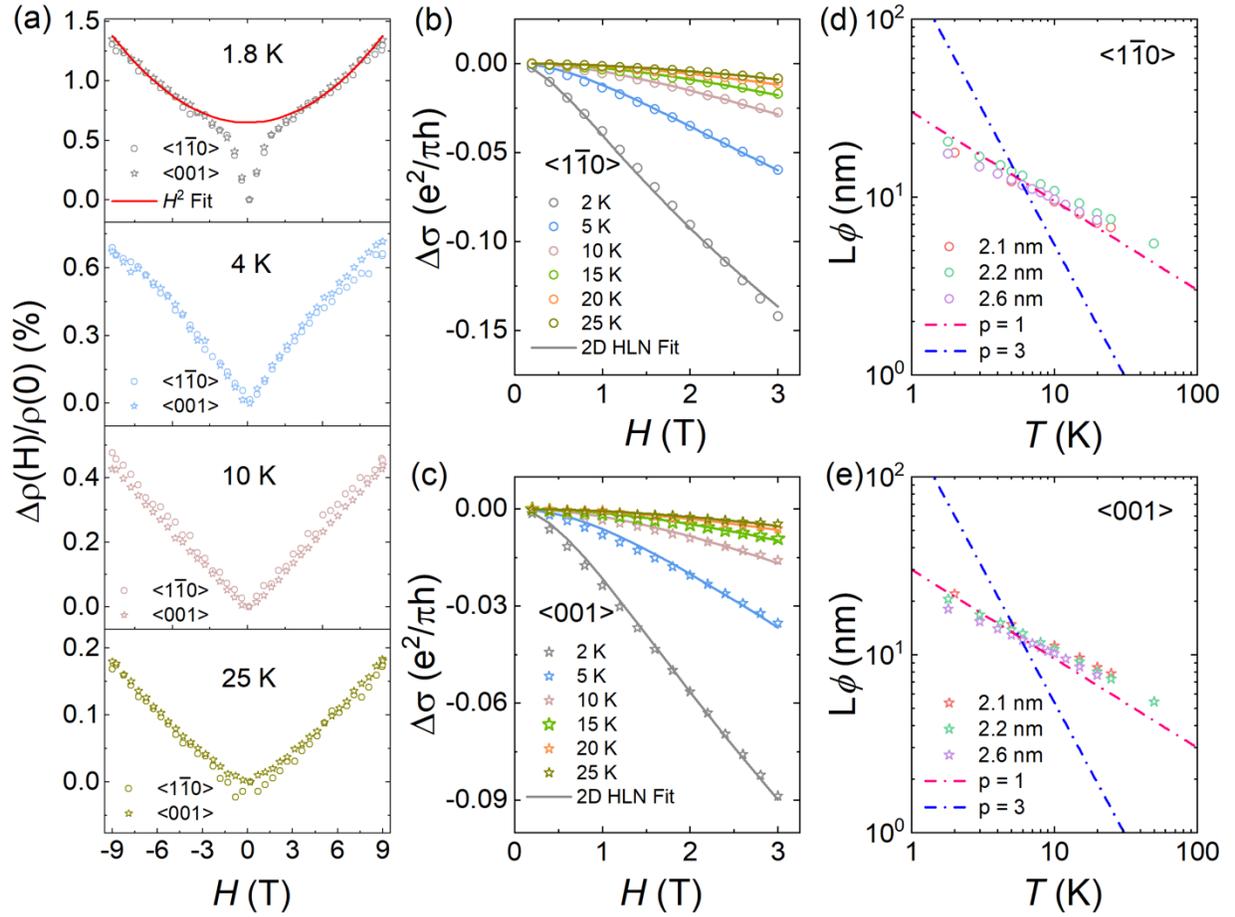

**Figure 3:** (a) Magnetoresistance in a perpendicular magnetic field for the 2.1 nm thick film at various temperatures. The red line in the top panel is a fit to $H^2$ behavior. Magnetoconductance as a function of magnetic field for (b) $<1\bar{1}0>$ and (c) $<001>$ crystalline directions including their fits to the 2D HLN model. Temperature dependence of inelastic scattering length for (d) $<1\bar{1}0>$ and (e) $<001>$ for three different samples showing weak anti-localization. The pink and blue dashed lines are theoretical curves for $p = 1$ and $p = 3$ respectively.